\documentclass{ws-ijmpd}
\usepackage[super,compress]{cite}
\usepackage{aas_macros}
\def\sgrze {SGR\,0418+5729}
\def\sgrsw {Swift\,J1822.3--1606}

\begin{document}

\markboth{R.~Turolla and P.~Esposito}
{Low-magnetic-field magnetars}

%
\catchline{}{}{}{}{}
%

\title{LOW-MAGNETIC-FIELD MAGNETARS\footnote{This is an author-created, un-copyedited version of an article accepted for publication in the International Journal of Modern Physics D.}}

\author{ROBERTO TUROLLA}

\address{Dept. of Physics and Astronomy, University of Padova, via F. Marzolo 8, I-35131 Padova, Italy\\
MSSL, University College London, Holmbury St. Mary, Dorking, Surrey RH5 6NT, UK\\
E-mail: roberto.turolla@pd.infn.it}

\author{PAOLO ESPOSITO }

\address{INAF -- IASF Milano, via E. Bassini 15, I-20133 Milano, Italy\\
E-mail: paoloesp@iasf-milano.inaf.it}

\maketitle

\begin{history}
\received{3 March 2013}
\accepted{5 March 2013}
\end{history}

\begin{abstract}
It is now widely accepted that soft gamma repeaters and anomalous X-ray pulsars are the 
observational manifestations of magnetars, i.e. sources powered by their own magnetic energy. This view 
was supported by the fact that these `magnetar candidates' exhibited, without exception, a surface dipole 
magnetic field (as inferred from the spin-down rate) in excess of the electron critical field ($\simeq$$4.4\times 10^{13}\ {\mathrm G}$). The recent discovery of fully-qualified magnetars, \sgrze\ and \sgrsw, 
with dipole magnetic field well in the range of 
ordinary radio pulsars posed a challenge to the standard picture, showing that a very strong field is not
necessary for the onset of magnetar activity (chiefly bursts and outbursts). Here we summarize the 
observational status of the low-magnetic-field magnetars and discuss their properties in the context of the 
mainstream magnetar model and its main alternatives.
\end{abstract}


\ccode{PACS numbers: 97.60.Jd, 97.60.Gb}

\section{Introduction}\label{intro}

A newly discovered and seemingly isolated neutron star is
(observationally) classified as a \emph{magnetar}\cite{paczynski92,duncan92,thompson95,thompson96,tlk02} 
(a source powered by magnetic energy)
when it complies with at least three of the following
requirements: comparatively long spin period \mbox{($P\sim 1$--12 s)};
large spin-down rate ($\dot P\gtrsim 10^{-12}$ s
s$^{-1}$); relatively high and variable persistent luminosity
($L_{\mathrm{X}}\sim 10^{32}$--$10^{36}$ erg s$^{-1}$);  emission
of powerful short X/$\gamma$-ray bursts (spikes of $\sim$0.1--10 s duration, $L_{\mathrm{X}}\sim
10^{34}$--$10^{47}$ erg s$^{-1}$ at the peak).

Just a few years ago, these criteria were thought to be equivalent
to that of a surface magnetic field of $10^{13}$--$10^{15}$ G. The
`magnetar candidates', in fact, comprise two classes of
sources, the soft gamma repeaters (SGRs) and the anomalous X-ray
pulsars (AXPs), which, with no exception at that time, exhibited a
surface dipole field (as derived from the spin down measure) in
excess of the quantum critical field, $B_{\mathrm Q}\simeq
4.4\times 10^{13}$ G.\cite{mereghetti08} Despite SGRs and AXPs are far from being a
homogeneous class, in particular their magnetic field spans nearly
two orders of magnitude, their observational behavior has been
assumed as the template for (active) magnetars, to the point that
 the terms SGR/AXP and magnetar are often used
interchangeably. This, actually, reflects the original definition
of a magnetar as a neutron star which is powered by its (large) magnetic
field.\cite{thompson93} In this respect, it is important to notice that a super-strong
magnetic field is not per se a sufficient condition for triggering
SGR/AXP-like activity. This is testified by the existence of neutron-star
sources, for instance most of the so-called high-$B$ radio pulsars
(HBPSRs; e.g. Refs.\,\refcite{kaspi05}, \refcite{kaspi10}), and  some of the thermally
emitting isolated neutron stars (XDINSs; e.g. Ref.\,\refcite{turolla09}), with
surface magnetic fields comparable to those of SGRs/AXPs but
having substantially different properties and not showing any
bursting/outbursting activity over the $\sim$10--20 yr time span
during which they have been observed.

More recently the `supercritical $B$' paradigm for the onset of
magnetar activity has been challenged the discovery of
full-fledged magnetars, SGR\,0418+5729 and Swift\,J1822.3--1606,
\cite{rea10,turolla11,rea13,livingstone11,rie12,scholz12} with a dipole magnetic
field well in the range of ordinary radio pulsars. Here we
consider whether the canonical magnetar model still holds in the
light of these facts or other models can better explain the
observed phenomenology. Section\,\ref{measB} is a brief 
recap of how the magnetic field of isolated pulsars is routinely estimated 
within the standard  magneto-dipole braking framework. 
Section\,\ref{observations} reviews the observational results on SGR\,0418+5729 and 
Swift\,J1822.3--1606. Sections \ref{theory} and \ref{alternative} deal 
with the theoretical context. Section \ref{conclusion} concludes this 
review with remarks on the definition of magnetar.

\section{Measuring the magnetic field of an isolated pulsar}\label{measB}

The surface dipole magnetic field of a non-accreting pulsar is
usually estimated by equating the rate of rotational kinetic
energy loss to the power radiated by the rotating dipole. In this
way, at the neutron-star magnetic equator,\footnote{See
Ref.\,\refcite{spitkovsky06} for a more accurate expression.} $B_p =
(3Ic^3 \dot{P}P/(8\pi^2 R^6\sin^2\alpha))^{1/2} \simeq
3.2\times10^{19}(P \dot{P}/\sin^2\alpha)^{1/2}~{\rm G}$, where the
period $P$ is measured in s, the period derivative $\dot{P}$ in s
s$^{-1}$, $\alpha$ is the angle between the magnetic moment and
the spin axis, and $R=10^6$ cm and $I=10^{45}$ g cm$^2$ are the
neutron-star radius and moment of inertia.

This inference is model-dependent, $R$ and $I$ are uncertain, and
$\alpha$ is generally unknown  (it is usually assumed, for
simplicity, to be 90$^\circ$), but no direct measurements of the
magnetic field strength are available for isolated
pulsars.\footnote{With the possible exception of 1E\,1207.4--5209,
the spectral features of which are often interpreted as electron
cyclotron lines.\cite{bignami03,suleimanov10,suleimanov12}} For
this reason, one has necessary to rely upon the approximate value
from the $B_p\propto(P \dot{P})^{1/2}$ formula and thus on
measurements of the rotational parameters.

The pulsar period and period derivative can be measured with good
precision by means of phase-coherent timing (see e.g.
Refs.\,\refcite{phinney92},\,\refcite{kaspi01},\,\refcite{dallosso03}) during extensive
observational campaigns. The basic idea is that in a reference
frame that does not accelerate with respect to an isolated pulsar
(to a very good approximation, the centre of mass of the Solar
System is an inertial reference frame), the time evolution of the
pulse phase $\phi(t)$ is, in general, expected to be well
described by the Taylor expansion
$\phi(t)=\phi(t_0)+\nu(t-t_0)+\frac{1}{2}\dot{\nu}(t-t_0)^2+\frac{1}{6}\ddot{\nu}(t-t_0)^3+\ldots$,
where $\nu=1/P$ is the pulse frequency and $t_0$ is a reference
epoch. The pulsar period and its successive derivatives are thus
obtained by the fit of the expansion to the observed data.

\section{Low-Magnetic-Field Magnetars: Observations}\label{observations}
\subsection{SGR\,0418+5729}
As for many recently discovered magnetars,\cite{rea11} the
existence of SGR\,0418+5729 was revealed by its emission of short
bursts. This occurred on 2009 June 5, when a couple of events
triggered the monitors for hard X-ray transients aboard
\emph{Fermi}, \emph{Koronas-Foton}, and
\emph{Swift}.\cite{vanderhorst10} These bursts were comparatively
faint ($L_{\mathrm{X}}\approx 10^{38}$--$10^{39}$ erg s$^{-1}$ in
the band 8--200 keV, for a distance\footnote{The low absorption
toward SGR\,0418+5729 and its direction are consistent with it
belonging to the Perseus Arm of the Galaxy, suggesting a distance
of $\sim$2 kpc.\cite{vanderhorst10,esposito10,durant11}} $d=2$
kpc) but otherwise unremarkable, with spectra well described by an
optically-thin thermal bremsstrahlung with temperature $kT\sim
20$--30 keV and duration of $\sim$40--80 ms.\cite{vanderhorst10}
A third possible (weaker) burst, again on June 5, was found by an
off-line inspection of the \emph{Fermi}/GBM data, but it was not
confirmed by the simultaneous \emph{Swift}/BAT observation; a
search of the Interplanetary Network\footnote{See
\mbox{http://heasarc.nasa.gov/W3Browse/all/ipngrb.html}.} events in
the period 1990--2009 did not reveal any past activity clearly
associated to SGR\,0418+5729.\cite{vanderhorst10}

In the following few days, follow-up observations in the soft
X-ray band (1--10 keV) carried out with \emph{Swift},
\emph{Chandra} and \emph{RossiXTE} unveiled the existence of a
previously unknown bright source, with an observed flux of a few
\mbox{$10^{-11}$ erg cm$^{-2}$ s$^{-1}$} (previous upper limits, based on
\emph{ROSAT} All-Sky Survey data, were of the order of $10^{-12}$
erg cm$^{-2}$ s$^{-1}$).\cite{vanderhorst10,esposito10} These
observations also made it possible to measure the source spin
period, $P\simeq9.1$ s.\cite{vanderhorst10} The results of the
first $\sim$5 months of monitoring of SGR\,0418+5729 are described
in Ref.\,\refcite{esposito10}. In each observation the X-ray
spectrum could be well described by a single- or two-blackbody
model (depending on the count statistics of the data
sets).\cite{esposito11,turolla11} During this stretch of time, the
source persistent X-ray emission faded by a factor of $\sim$10,
with a clear softening with time. The surprising fact was that,
despite the dense monitoring and the long time-span, no slowing of
the pulsar rotation could be detected, with a 3$\sigma$ upper
limit of \mbox{$1.1\times10^{-13}$ s s$^{-1}$}, translating into an upper
limit on the surface dipole magnetic field strength (see
Sec.\,\ref{measB}) of $3\times10^{13}$ G.\cite{esposito10}

This already made SGR\,0418+5729 the magnetar with the lowest
inferred dipole magnetic field.\footnote{See the McGill Pulsar
Group AXP/SGR catalogue at the web page
\mbox{http://www.physics.mcgill.ca/~pulsar/magnetar/main.html}.}
Although indeed small for a magnetar, such a value was not
abnormal, the limit being comparable to the strength of
$6\times10^{13}$ G estimated for the dipole magnetic field of the
AXP 1E\,2259+586. It is interesting to note that the limit on the
period derivative implied also a limit on the spin-down flux of
$\dot{E}/(4\pi d^2)=\pi I\dot{P}/(P^{3}d^{2})<2\times10^{-14}$ erg
cm$^{-2}$ s$^{-1}$, showing that the persistent luminosity of the
source during the monitoring could not be rotation-powered.

After a period during which the direction of the source was not
accessible to the X-ray spacecrafts due to Sun constraints, a new
observational campaign, mainly aimed at achieving a measurement of
the period derivative, was undertaken in 2010 July through
September using \emph{Swift}, \emph{Chandra}, and
\emph{XMM-Newton}. Again, the period derivative of SGR\,0418+5729
eluded detection. This time, however, the limit on the period
derivative, obtained through coherent timing over a base line of
approximately 500 days, was substantially deviating from what
could be expected for a magnetar: $\dot{P}<10^{-15}$ s s$^{-1}$,
corresponding to $B_p<7.5\times10^{12}$ G (90\% confidence
level).\cite{rea10} Not only, in fact, this value was
significantly below the electron quantum magnetic field
$B_{\mathrm{Q}}\equiv m^2_e c^2/(\hbar e)\simeq
4.4\times10^{13}$ G (a value that, albeit lacking direct physical
implications for pulsars, was traditionally considered to be the
divide between ordinary pulsars and magnetars), but magnetar-like
activity (bursting/outbursting behavior in particular) had never been observed
before in objects with a magnetic field this low (also the
magnetar/pulsar PSR\,J1846--0258\footnote{PSR\,1846--0258,\cite{gotthelf00} 
at the centre of  the supernova remnant Kes 75, is a young  and energetic 
0.3-s pulsar (characteristic age 
smaller than 1 kyr, spin-down power of $\sim$$8\times10^{36}$ erg s$^{-1}$) 
powering a bright pulsar wind nebula; no radio emission has been detected from
this pulsar. With an X-ray luminosity well below 
its spin-down power, it was considered an ordinary rotation-powered pulsar until 
in 2006 it underwent a magnetar-like outburst accompanied by the emission of 
several short bursts.\cite{gavriil08} The classification of this source is now debated 
and it is often indicated as a `hybrid' or `transitional' object and sometimes counted 
among magnetars.} has an inferred surface dipole magnetic field of $4.9\times10^{13}$ G, in excess
of $B_{\mathrm{Q}}$).

\begin{figure}[pb]
\centerline{\psfig{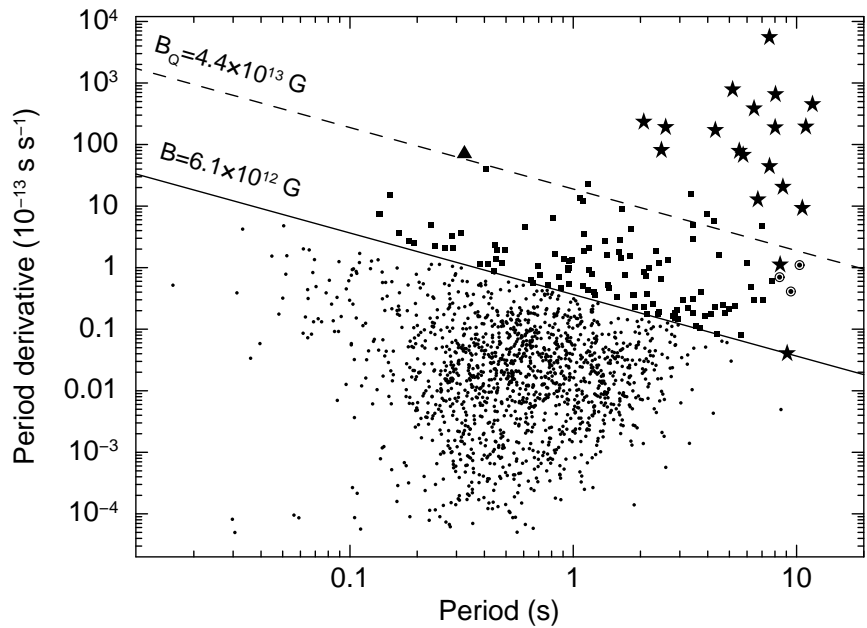}}
\vspace*{8pt}
\caption{$P$--$\dot P$ diagram for non-recycled pulsars (data are 
from Ref.\,31). Points represent normal radio pulsars, squares normal 
radio pulsars with a magnetic field larger than that measured for 
SGR\,0418+5729, stars are the magnetars, the triangle is the magnetar-like 
pulsars PSR\,J1846--0258, and the circled dots are the X-ray dim isolated 
neutron stars. The solid line marks the dipole magnetic field inferred for 
SGR\,0418+5729. The value of the electron quantum magnetic field is also 
shown (dashed line). \label{ppdot}}
\end{figure}

Since during the Summer 2010 the source observed flux was of
approximately \mbox{(1--$2)\times10^{-13}$ erg cm$^{-2}$ s$^{-1}$}, the spectrum
further softened and there was no sign that the flux decline had
stopped, the monitoring of SGR\,0418+5729 was extended using only
the high-effective-area detectors on board \emph{Chandra} (three
pointings between 2010 November and 2011 November) and
\emph{XMM-Newton} (four pointings between 2011 March and 2012
August). \cite{rea13} In the time span covered by these
observations, the source flux apparently settled at about
$(1$--$2)\times10^{-14}$ erg cm$^{-2}$ s$^{-1}$, which could be
the typical quiescent emission level of SGR\,0418+5729 (3 orders
of magnitude below that measured at the time of the discovery,
see Sec.\,\ref{theory} for the long-term X-ray light curve).
Finally, after more than 3 years of monitoring, the measurement of
the source spin-down rate was achieved, at a 3.5$\sigma$
significance level (from a coherent timing analysis of all the
X-ray data spanning $\sim$1200 days; see
Refs.\,\refcite{esposito10}, \refcite{rea10}, and \refcite{rea13}
for more details): \mbox{$(4\pm1)\times10^{-15}$ s s$^{-1}$},
corresponding to $\sim$$6.1\times10^{12}$ G. The value also
translates into a characteristic age $\tau_c=P/(2\dot{P})\simeq36$
Myrs and spin-down power
$\dot{E}=4\pi^2I\dot{P}/P^3\simeq2\times10^{29}$ erg s$^{-1}$
(showing that the X-ray luminosity of SGR\,0418+5729 cannot be
rotation powered not even at the low level observed in the
2011--2012 campaign). Figure\,\ref{ppdot} shows the position of
SGR\,0418+5729 in the  $P$--$\dot P$ diagram for pulsars, while
Table\,\ref{tab1} summarizes its principal characteristics.

\begin{table}
\tbl{Main characteristics of SGR\,0418+5729 and
Swift\,J1822.3--1606.} {\begin{tabular}{@{}lcc@{}} \toprule
 & SGR\,0418+5729 & Swift\,J1822.3--1606$^{\text a}$\\
\colrule
RA (J2000) & $\rm04^h\,18^m\,33.87^s$ & $\rm18^h\,22^m\,18.06^s$ \\
Dec. (J2000) & $+57^\circ\,32'\,22.91''$ & $-16^\circ\,04'\,25.55''$ \\
Distance (kpc) & 2 & 1.6 \\
Period (s) & 9.07838822(5) & 8.43771984(4)\\
Period derivative (s s$^{-1}$) & $4(1)\times10^{-15}$ & $1.71(7)\times10^{-13}$\\
Reference epoch (MJD) & 54993.0 & 55761.0\\
Validity range & 54993--56164& 55759--56161\\
Surface dipole field (G) &  $6.1\times10^{12}$ & $3.8\times10^{13}$\\
Spin-down power (erg s$^{-1}$) & $2.1\times10^{29}$ & $1.1\times10^{31}$ \\
Characteristic age (Myr) & 36 & 0.8 \\
Luminosity (0.5--10 keV, erg s$^{-1}$)& $8\times10^{30}$---$1.6\times10^{34}$$^{\text b}$ & $3\times10^{32}$---$9\times10^{35}$$^{\text c}$ \\
\botrule
\end{tabular}}
\begin{tabnote}
Values in parentheses are 1$\sigma$ uncertainties in the least
significant digit quoted. The 95\%-confidence positional error
radius is 0.4$''$ for SGR\,0418+5729 and 0.7$''$ for
Swift\,J1822.3--1606. $^{\text a}$ For Swift\,J1822.3--1606
different values of the timing parameters can be found in
literature (see main text and references therein); here we give
the $P$--$\dot{P}$--$\ddot{P}$ coherent solution by
Ref.\,\refcite{scholz12}. $^{\text b}$ Minimum and maximum values,
derived from the fluxes measured at the end and the start,
respectively, of the the 2009 June--2012 August campaign
(Refs.\,\refcite{esposito10}, \refcite{rea13}). $^{\text c}$
Minimum and maximum values, corresponding to the 1993 September 14--15
\emph{ROSAT} observation and to a \emph{Swift} observation carried
out on 2011 July 15 (at the start of the outburst), respectively.
\end{tabnote}
\label{tab1}
\end{table}

\subsubsection{Multi-wavelength observations}

Until recently, magnetars were believed to be emitting essentially
in the soft X-ray energy range, briefly reaching the soft $\gamma$-ray
energies during their bursts. Thanks to deep surveys and the
availability of new instruments, nowadays several magnetars have
been observed in the
radio\cite{camilo06,camilo07,levin10,anderson12}, optical and
infrared,\cite{testa08,rct10,mignani11,tendulkar12} and soft
$\gamma$-ray\cite{kuiper04,kuiper06,gotz06,rea09,kaneko10} (as
persistent sources) wavebands. Most counterparts were identified
during outbursts,\cite{rea11} because during active periods the
emission level is in general enhanced at all wavelengths and the
rapid flux variability helps in sorting out the true counterpart
when there are many candidates.

On 2009 June 19, ten days after the onset of the outburst, the
field of SGR\,0418+5729 was observed by the Green Bank Telescope
at 820 MHz but no source was detected, with an limit on the flux
density of $<$0.5 mJy (for a duty cycle of 20\%).\cite{lorimer09}
At millimeter wavelengths, five observations were carried out at
the Plateau de Bure Interferometer between 2011 June and July;
again, no signal from SGR\,0418+5729 was detected, with a limit of
0.24 mJy at 1.8 mm (167 GHz).\cite{rea13} The source eluded
detection so far also in the optical and infrared bands. The
magnitude limits  from observations performed by ground-based
instruments (Ref.\,\refcite{esposito10} and references therein)
are $K_s>22.9$ (Palomar Hale Telescope, on 2009 August 2), $r>24$
(William Herschel Telescope, on 2009 August 16), and $i>25.1$
(Gran Telescopio Canarias, on 2009 September 15). Finally, an
observation with the \emph{Hubble Space Telescope} on 2010 October
19 yielded magnitude limits of 28.6 in the visible (5921 \AA)
and 27.4 in the infrared (11534 \AA).\cite{durant11}

\subsection{Swift\,J1822.3--1606}

Swift\,J1822.3--1606 was discovered on 2011 July 14,\footnote{The
Summer of 2011 was a bountiful one for magnetar enthusiasts: on 2011 August 7, less 
than one month after the discovery of \sgrsw, another new magnetar 
was discovered, Swift\,J1834.9--0846.\cite{kargaltsev12,etr13}}
when it emitted several magnetar-like bursts that were detected by the
\emph{Swift}/BAT (see Refs.\,\refcite{livingstone11},
\refcite{rie12} and references therein). Soon after, follow-up
observations with \emph{Swift} and \emph{RossiXTE} led to the
detection of a bright X-ray counterpart (flux of
$\sim$$2\times10^{-10}$ erg cm$^{-2}$ s$^{-1}$ in the first days
after the onset of the outburst, with a spectrum well described by
either a blackbody plus power law or a two-blackbody model),
pulsating at \mbox{$\sim$8.4 s}. The spin period, together with the short
bursts and the spectral properties, as well as the lack of an
optical or infrared counterpart, confirmed the magnetar nature of the
source.

Although previously unnoticed, the source was already present in
two \emph{ROSAT} X-ray catalogues, having been serendipitously
detected in 1993 September 14--15 at a flux level of $\sim$$4\times10^{-14}$ 
erg cm$^{-2}$ s$^{-1}$.\cite{rie12,scholz12} A
distance of $\sim$1.6 kpc has been proposed on the basis of a
possible association, supported by the similar absorption columns,
between the source and the H\textsc{ii} region M17.\cite{scholz12}
Adopting this value, the 1993 luminosity was roughly
$3\times10^{32}$ erg s$^{-1}$, while it was $\sim$$9\times10^{35}$
erg s$^{-1}$ in 2011 July (0.5--10 keV). After its (re)discovery,
Swift\,J1822.3--1606 was intensely monitored between 2011 July and
2012 August using \emph{Swift}, \emph{RossiXTE}, \emph{Chandra}
and \emph{XMM-Newton}.\cite{livingstone11,rie12,scholz12} In this
period, during which the source remained moderately
burst-active,\cite{scholz12} the luminosity declined from the
discovery value to $\sim$$10^{33}$ erg s$^{-1}$ (see
Refs.\,\refcite{rie12}, \refcite{scholz12}  and Sec.\,\ref{theory} for 
the long-term X-ray curve).

Based on the data collected between 2011 July and October,
Ref.\,\refcite{livingstone11} proposed a period derivative of
$\sim$$2.5\times10^{-13}$ s s$^{-1}$, corresponding to a surface
dipole magnetic field of $\sim$$4.7\times10^{13}$ G, a value lower
than that of 1E\,2259+586 and comparable to that of the
magnetar/pulsar PSR\,J1846--0258. Using a partially different set
of observations covering the period from 2011 July to 2012 April,
Ref.\,\refcite{rie12} revised the spin-down rate and the magnetic
field to even lower values of  $\sim$$8\times10^{-14}$ s s$^{-1}$
and $\sim$$2.7\times10^{13}$ G. Recently, Ref.\,\refcite{scholz12}
proposed new coherent timing solutions using the data of
Ref.\,\refcite{livingstone11} plus several new observations that
extended the base line to 2012 August (spanning $\sim$400 days).
These solutions, which differ one from another for the number of
frequency derivatives considered in the Taylor series used to
fit the pulse phase evolution (see Sec.\,\ref{measB}), yield
spin-down rates between approximately \mbox{$0.7\times10^{-13}$ s
s$^{-1}$} and $3.1\times10^{-13}$ s s$^{-1}$, implying a magnetic
field between $2.4\times10^{13}$ G and $5.1\times10^{13}$ G. The
reason for these discrepancies and the multiple possible solutions
is likely the `timing noise' (e.g.  Ref.\,\refcite{hobbs10}), a
common phenomenon in young neutron stars and magnetars in
particular.\cite{esposito11} If so, the timing properties of Swift\,J1822.3--1606
probably will need longer time to be properly characterized.
Nevertheless, any of the values proposed so far in literature would
make Swift\,J1822.3--1606 the magnetar with the second lowest
inferred magnetic field after SGR\,0418+5729. Finally, also for
this source the timing analysis dismisses the possibility that the
X-ray emission could be powered to a substantial extent by the
loss of rotational energy (see Table\,\ref{tab1}).

\subsubsection{Multi-wavelength observations}

Significant efforts have been devoted to detect emission in energy
bands other than X-ray from Swift\,J1822.3--1606 but so far, as
for SGR\,0418+5729, without success. At radio frequencies, the
field of Swift\,J1822.3--1606 was pointed four times between 2011
July and October with the Green Bank Telescope, but no pulsed
emission was detected down to a flux density of 0.05--0.06
mJy.\cite{rie12}

An observation with the $z$ filter (9694 \AA) was carried out at
the Gran Telescopio Canarias on 2011 July 21, resulting in a
limiting magnitude $z>22.4$.\cite{rie12} The field was also
serendipitously imaged on 2006 May 3 by the UK Infrared Telescope;
the analysis of this pre-2011-outburst observation yielded
infrared limiting magnitudes $J>19.3$, $H>18.3$, and $K>17.3$ (see
Ref.\,\refcite{rie12} and references therein for details on this
and other optical and infrared observations).

\section{Low-Magnetic-Field Magnetars In The `Standard
Model'}\label{theory}

As discussed in the previous section, overall the observed
properties of \sgrze\ and \sgrsw\ are not dissimilar from those of
other known (transient) magnetar candidates.\cite{mereghetti08,rea11} 
The blatant difference is in the estimated strength of the dipole field $B_p$,
which is well in the ordinary pulsar range, especially for \sgrze.
There is, moreover, a further point which sets the two low-$B$
sources apart: their long characteristic age, $\tau_c\gtrsim 10^6$ yr,
as compared to $\approx$$10^3$--$10^4$ yr for the other magnetar
candidates. The latter is clearly a consequence of the former, and
both reflect the smallness of the period derivative.

Albeit the characteristic age is well known for providing a poor
estimate of the neutron star true age, the very large values of
$\tau_c$ may suggest that \sgrze\ and \sgrsw\ are old objects. The
small number of detected bursts (with comparatively low
energetics) and the low persistent luminosity in quiescence have
been taken as further hints that these might be worn-out
magnetars, approaching the end of their active life \cite{esposito10,rea10,
turolla11}. The `old magnetar' scenario sounds appealing since
it offers an interpretation of the low-$B$ sources within an
already established framework, validating the magnetar model also
for (surface) field strengths quite far away from those of
canonical SGRs/AXPs. At the same time, it rises a number of
crucial questions, chiefly how one can reconcile the low dipole
field with the huge magnetic stresses required to deform the crust
and produce the bursts/outbursts.

Actually, one should bear in mind that the ultimate powerhouse of
a (active) magnetar is its internal field, its toroidal component
in particular. So, it is possible that in low-$B$ magnetars, and
in other neutron star sources as well \cite{deluca12,shabaltas12,vigano12}, the
magnetic field is `hidden' in the star interior and only a
relatively weak dipolar component emerges. Still, if \sgrze\ and
\sgrsw\ were born with a magnetar-like surface field, $B_p$ must
have decayed by a factor $\approx$10--100 to match the current
values. Roughly the same reduction is expected in the internal
field. Although the latter could initially be $\approx$10 times
higher than $B_p$ (at least locally), one may wonder if at late
times internal magnetic stresses are still strong enough to crack
the crust. A second and related question is if realistic models of
field decay in magnetars can account for the observed rotational
properties (period, period derivative) and quiescent luminosity of
\sgrze\ and \sgrsw. This also directly bears to the true age of
the sources which are most probably (much) younger than their
characteristic age, estimated assuming a non-decaying field.

\subsection{Magneto-rotational Evolution}\label{magevol}

The more general configuration for the internal field in a neutron star will
be that produced by the superposition of current systems in the
core and the crust. As stressed by Ref.\,\refcite{pg07}, the relative
contribution of the core/crustal fields is likely different in
different types of neutron stars. In old radio pulsars, where no field decay
is observed, the long-lived core component may dominate, while a
sizable, more volatile crustal field is probably present in
magnetars, for which substantial field decay over a timescale
$\approx$$10^3$--$10^5$ yr is expected (see e.g. Ref.\,\refcite{goldreich92}).

In magnetars, because of the lesser role of ambipolar diffusion in
the core,\cite{glampedakis11} the decay/evolution of the magnetic field
is likely to take place in the crust and is governed by Hall/Ohmic
diffusion. The relative importance of these two mechanisms is
strongly density- and temperature-dependent. Thus, any
self-consistent study of the magnetic field evolution must be
coupled to a detailed modeling of the neutron star thermal
evolution, and conversely. This basically means that the induction
equation for $\mathbf{B}$ must be solved together with the
cooling, a quite challenging numerical task.

In recent years, much efforts have been devoted to this problem
\cite{pg07,aguilera08,pons09}. The state-of-the-art numerical
code is that of Ref.\,\refcite{pons09}, which features full
magneto-thermal coupling and includes all realistic microphysics.
However, owing to numerical difficulties in treating the Hall
term, the models in Ref.\,\refcite{pons09} include only Ohmic diffusion. This can be a
limitation because the Hall drift likely drastically affects the
very early evolution of ultra-magnetized neutron stars with surface field
$B_p\gtrsim 10^{15}$~G. However, for lower $B_p$, still well
within the magnetar range, the effect of the Hall drift is
expected to introduce at the most quantitative changes (a somewhat
faster dissipation) with respect to the purely Ohmic picture. Very
recently a code capable of properly handling the Hall term has
been presented \cite{vpm12}, but applications to neutron star
magneto-thermal evolution are still to come.

\begin{figure}[pb]
\centerline{\psfig{file=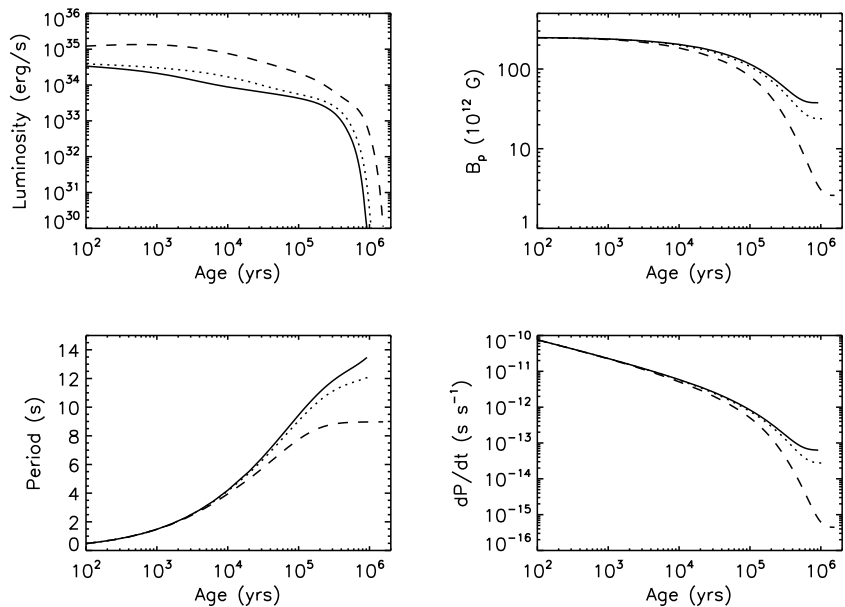,angle=0,width=12cm}}
\vspace*{8pt}
\caption{From top left to bottom right: time evolution of $L_{\mathrm{X}}$, $B_p$, $P$, $\dot P$
for \sgrze. The three curves in each panel refer to  $B_{tor}(t=0)=0$ (solid lines), $B_{tor}(t=0)=4\times
10^{14}$~G (dotted lines) and $B_{tor}(t=0)=4\times 10^{16}$~G
(dashed lines). Figure taken from Ref.\,12. \label{magevol-0418}}
\end{figure}

\begin{figure}[pb]
\centerline{\psfig{file=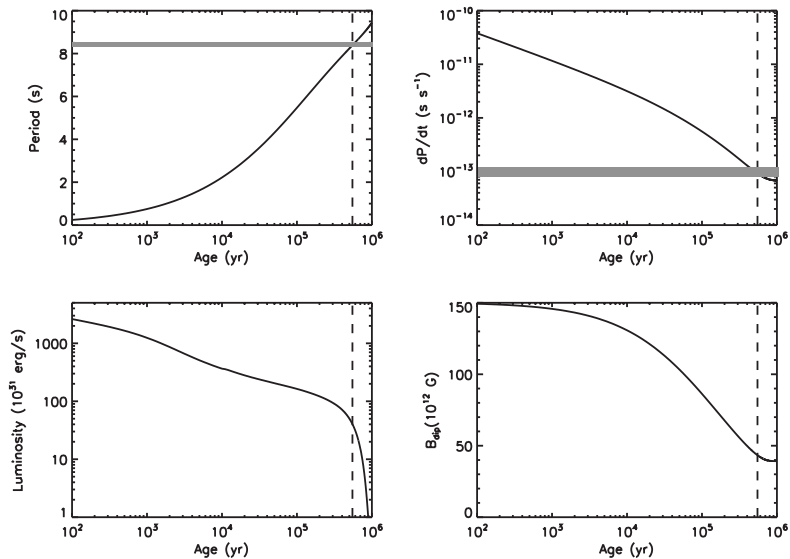,angle=0,width=12cm}}
\vspace*{8pt}
\caption{From top left to bottom right: time
evolution of $P$, $\dot P$, $L_{\mathrm{X}}$ and $B_p$ for \sgrsw. The dashed
vertical line marks the estimated age of the source; the gray
strips in the first two panels show the observed values of $P$ and
$\dot P$ with their uncertainties. The model is for
$B_{tor}(t=0)=5\times 10^{15}$~G. Figure taken from
Ref.\,15. \label{magevol-1822}}
\end{figure}

In order to explore if, and to which extent, the magneto-thermal
evolution of (initially) highly magnetic neutron stars can lead to objects
with properties compatible with those of \sgrze\ and \sgrsw,
Refs.\,\refcite{turolla11} and \refcite{rie12}
performed a number of runs using the code of Ref.\,\refcite{pons09}. 
Results for the two sources are shown in Figs.
\ref{magevol-0418} and \ref{magevol-1822}. The main outcome is
that magnetic field decay in an initially ultra-magnetized neutron star,
$B_p(t=0)\sim 2\times 10^{14}$ G, can reproduce the observed $P$,
$\dot P$, $B_p$ and $L_{\mathrm{X}}$ in \sgrze\ and \sgrsw, for an age $\sim$1 Myr
and $\sim$0.5 Myr, respectively, provided that the initial
internal toroidal field $B_{tor}(t=0)$, the key parameter, is high
enough, $\sim$$10^{16}$ G in the former and $\sim$$5\times 10^{15}$
G in the latter. Evolutionary calculations confirm that these are
old sources, as expected, although the true age is shorter than
$\tau_c$, the difference being more than one order of magnitude
for \sgrze.

\subsection{Active, Till the End}

Recently Ref.\,\refcite{perna11} used the magnetic evolution
code of Ref.\,\refcite{pg07} together with the cooling
models by Ref.\,\refcite{pons09} to compute the magnetic
stresses acting on the neutron-star crust at different times. Their baseline
model has $B_p(t=0)=8\times 10^{14}$~G and
$B_{tor}(t=0)=10^{15}$~G. They found that the occurrence of
crustal fractures (and hence of bursts/outbursts) is not
restricted to the early neutron star life, during which the surface field is
ultra-strong, but can extend to late phases ($\approx$$10^5$--$10^6$ 
yr; see their figure 2). Both the energetic and
the recurrence time of the events evolve as the star ages. For
`old' magnetars about $50\%$ crustal fractures release $\approx$$10^{41}$~erg 
and the waiting time between two successive events is
$\approx$1--10~yr. They also made a longer run with a model with
$B_p(t=0)=2\times 10^{14}$~G and $B_{tor}(t=0)=10^{15}$~G, for
which the event rate is about a factor 10 smaller.

The models which successfully reproduce the properties of \sgrze\
and \sgrsw\ have $B_p(t=0)$ very close to this latter
configuration, while $B_{tor}(t=0)$ is larger. On the basis of
this, although no dedicated simulations have been performed,
Refs.\,\refcite{turolla11} and \refcite{rie12}
concluded that the two low-$B$ magnetars can become burst-active
despite their age, with an expected (present) event rate similar
to what predicted by the second model of Ref.\,\refcite{perna11}, i.e.
$\approx$$0.01$--$0.1\, \mathrm{yr}^{-1}$.

\subsection{Outburst Decay}
\begin{figure}[pb]
\centerline{\psfig{file=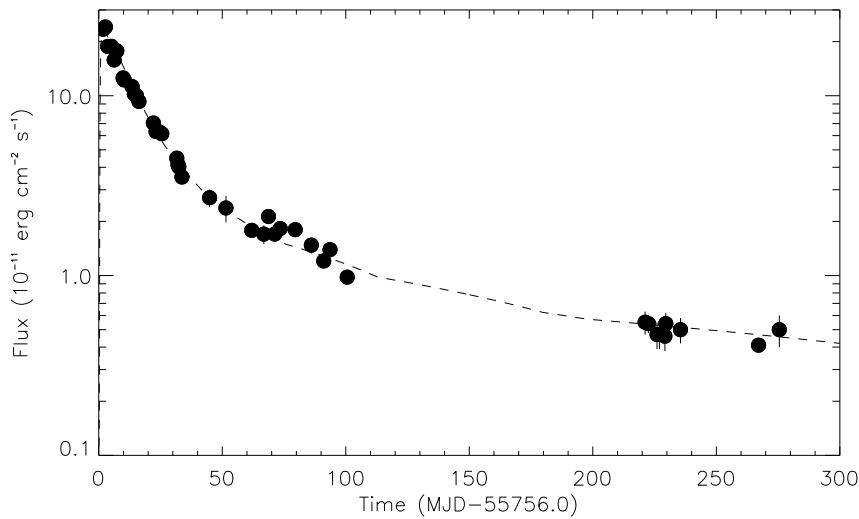,angle=0,width=12cm}}
\vspace*{8pt}
\caption{Flux (in the 1--10 keV band) decay following the outburst of \sgrsw. The trigger time of the first burst detected
from this source was MJD 55756.533.
The filled circles mark the observations and the 
curve the best model. Figure adapted from Ref.\,15. \label{outb-1822}}
\end{figure}

Outbursts, a distinctive trait of `transient' magnetars which
\sgrze\ and \sgrsw\ belong to,\cite{rea11} are characterized by a sudden
increase in flux (up to a factor $\approx$1000 over the quiescent
level) followed by a slow decay which lasts months/years. In many
sources the spectrum during the outburst is thermal (modeled by
one or two blackbodies with $kT\sim 0.3$--0.9 keV) and typically
softens during the decay. The radiation radius of the emitting
area(s) is small $\approx$0.1--1 km and usually decreases in
time. This has been interpreted as due to some form of heat
deposition in a limited region of the star surface which then
cools and shrinks. Up to now, however, the heating mechanism has
not been unanimously assessed. A possibility is that energy is
injected deep in the crust, e.g. because of magnetic dissipation,
and then flows to the surface \cite{let02}. Alternatively,
heating may be produced by currents flowing in a twisted
magnetosphere as they hit the star \cite{beloborodov09}.

\begin{figure}[pb]
\centerline{\psfig{file=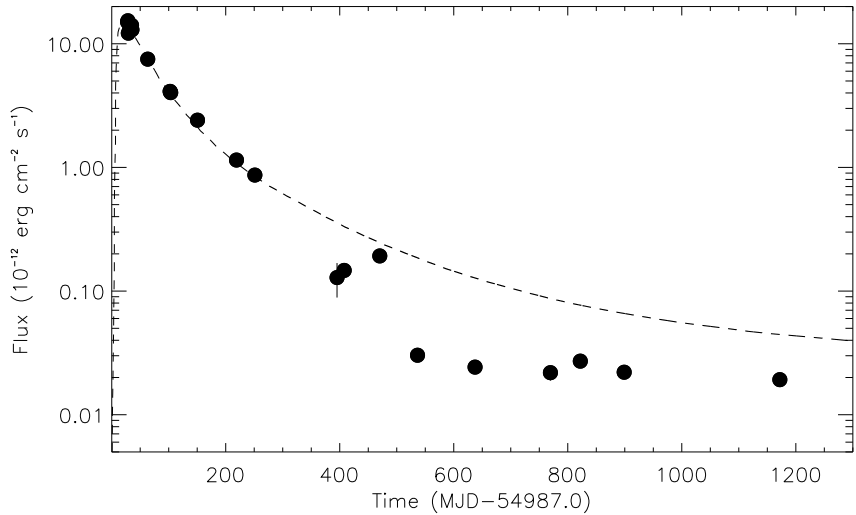,angle=0,width=12cm}}
\vspace*{8pt}
\caption{Flux (in the 0.5--10 keV band) decay 
following the outburst of \sgrze. The trigger time of the first burst detected
from this source is MJD 54987.862.
The filled circles mark the observations and the 
curve the model which best fits the early ($<$400~d) evolution.
Figure adapted from Ref.\,13. \label{outb-0418}}
\end{figure}

Very recently, Ref.\,\refcite{pons12} developed a
quantitative model for the outburst evolution by simulating the
thermal relaxation of the neutron star in response to an impulsive energy
injection in the star crust. Results were successfully applied by
Ref.\,\refcite{rie12} to fit the outburst decay in \sgrsw\
for the entire period covered by their observations, $\sim$250~d after
the first burst that led to the discovery of the source (see Fig.\,\ref{outb-1822}). The case of \sgrze, for
which a much longer time coverage is available ($\sim$1200~d),
is, however, much less conclusive in this respect \cite{rea13}
(see Fig.\,\ref{outb-0418}). The calculated flux in the 0.5--10 keV
band systematically underestimates the observed one at later times
($\gtrsim$400~d), when the luminosity suddenly drops and the hotter
blackbody (initially at $kT\sim 0.9$~keV) disappears leaving only
a cooler component at $\sim$0.3~keV. As discussed in Ref.\,\refcite{rea13}, 
because of the limited spatial extent of the twist
and hence of the low luminosity released by ohmic dissipation,
current heating is also unlikely to explain the observed flux in
\sgrze. Only a long term monitor will assess if \sgrze\ is indeed
peculiar, or also \sgrsw\ is bound to show the same behavior at
late times.

\section{Low-Magnetic-Field Magnetars: Alternative Scenarios}\label{alternative}

Although many indirect evidences accumulated in favor of the
magnetar paradigm (e.g. Refs.\,\refcite{thompson95},\,\refcite{vietri07}) and despite it
proved to be quite successful in explaining the observed properties of
SGRs/AXPs, including those of the low-$B$ sources, no conclusive,
direct measure of the surface magnetic field has been claimed as of yet. Contrary
to other classes of X-ray pulsars, in which it has been possible
to infer $B_p$ from the detection of (electron) cyclotron lines in
their spectra\cite{truemper78} (see e.g Ref.\,\refcite{heindl04} for a
review), up to now spin-down remains the only way for SGRs/AXPS,
with all the ensuing uncertainties (chiefly the fact that
spin-down is indeed due to magneto-dipole losses alone).

Since the cyclotron energy for a particle of charge $e$ and mass
$m$ is \mbox{$E_{B}\sim 11.6 (m_e/m)(B/10^{12}\, {\mathrm G})\
\mathrm{keV}$} (here $m_e$ is the electron mass), the electron line
falls well above the X-ray range for fields $B_p\gtrsim 10^{14}\
{\mathrm G}$, becoming inaccessible to observations. However, the
proton line, at $\sim$$0.63(B/10^{14}\, {\mathrm G})\ \mathrm{keV}$
is squarely in the X-ray band for magnetars. Proton cyclotron
lines in magnetar atmospheres have been extensively investigated
\cite{zane01,ho01,hlp03,ho04,potekhin10,shp10} and
searched for in virtually all available observations but escaped
unambiguous detection.\footnote{The presence of proton cyclotron
features has been reported in the spectra of some magnetar bursts,
e.g. Refs.\,\refcite{strohmayer00}, \refcite{ibrahim02}, but never assessed with certainty; in
these cases the derived magnetic field is of the same order of
that implied by spin-down.}

The lack of a smoking gun, definitely proving that magnetar
candidates host an ultra-magnetized neutron star, stimulated in
the past the investigation of alternative scenarios, which could
explain the observed characteristics of SGRs/AXPs without
resorting to ultra-strong magnetic fields. The recent discovery of
low-$B$ sources, with magneto-rotational properties similar to
those of standard radio pulsars, renewed this interest. A
long-standing competitor of the magnetar scenario has been the
fossil-disc model, according to which SGRs/AXPs harbor a neutron
star with standard magnetic field ($\approx$$10^{12}$--$10^{13}\
{\mathrm G}$) which is effectively spun down by the interaction
with a debris disc left in the parent supernova explosion or after
a common envelope phase, e.g. 
Refs.\,\refcite{vanparadijs95},\,\refcite{chatterjee00}, and \refcite{alpar01}.
According to Ref.\,\refcite{alpar11}, if \sgrze\ was born with a
period longer than 70 ms and a low dipolar field ($B_p\approx 10^{12}\
{\rm G}$), the torque exerted by a fall-back disc can spin down the
star to the present period in $\gtrsim$$10^5$ yr. In 
Ref.\,\refcite{truemper10} a somehow similar scenario in which SGRs/AXPs are
neutron stars with a low dipole field and super-strong multipolar
components powered by accretion from a fall-back disc was recently
proposed.

Models not invoking neutron stars at all have also been discussed.
Massive ($\sim$$1.3$--1.4 $M_\odot$) white dwarfs endowed with high
(on white-dwarfs standards) magnetic fields were suggested as possible
powerhouses\cite{malheiro12}. The basic idea is that, being a white dwarf
$\approx$1000 times bigger than a neutron star, at comparable mass, its
moment of inertia is $\approx 10^6$ times larger. This implies
that rotational energy losses can be large enough to explain the
observed X-ray luminosity in SGRs/AXPs ($\approx$$10^{32}$--$10^{36}$ erg s$^{-1}$) 
even for quite low values of
the period derivative. In addition, since $B_p\propto I^{1/2}/R^3\propto
1/R^2$, the dipolar magnetic field derived from spin-down is much
lower in a white dwarf than in a neutron star with the same $P$ and $\dot P$. The
inferred values of $B_p\sim 10^8$--$10^{10}\ {\mathrm G}$ are
somehow high, but still consistent with those observed in white dwarf. In
this scenario all the SGRs/AXPs activity (bursts, outbursts, giant
flares) is powered by the relief of mechanical stresses, driven by
gravity overcoming centrifugal forces as the white dwarf spins down. The
change in the star oblateness produces a decrease of the moment of
inertia and, in turn, a sudden increase of the spin frequency (a
glitch). The energy which can be tapped is $\approx$$10\%$ of the
available gravitational energy, $(GM_{\mathrm{WD}}^2/R_{\mathrm{WD}})(\Delta
R/R)\sim 2\times 10^{51}\vert\Delta P\vert/P)\ \mathrm{erg}$. In
order to accommodate the energetic of bursts and also of the
giant flares, which is several orders of magnitude larger,\cite{mereghetti08} the white-dwarf
model requires $\vert\Delta P\vert/P$ in the range 
$\approx$$10^{-7}$--$10^{-3}$. The fastest spinning SGRs/AXPs ($P\lesssim 5$ s) 
might pose a problem for the stability of white dwarfs.
However, according to recent calculations\cite{boshkayev13}, the
minimum period of a rotating, massive white dwarf could be as small as 0.3 s.

A more exotic scenario involves quark stars (see Ref.\,\refcite{ouyed11} 
and references therein). A neutron star could evolve 
into a quark star following an increase of its core density due to spin down or 
accretion. A quark nova explosion (releasing gravitational energy 
up to $\approx$$10^{53}$ erg, mostly carried away by neutrinos) would mark this 
transition and cause the ejection of the neutron star iron-rich outer layers. Depending 
on the quark star initial spin frequency, the degenerate debris material would either 
form a shell co-rotating with the star or be confined in a Keplerian ring (in the case 
of fast rotation). SGRs/AXPs are, then, the observational manifestation of these 
quark star--ring/shell systems.

Such quark stars are expected to be ultra-magnetized, because
color ferromagnetism in the quark matter can give rise to  
a surface magnetic field of $\approx$$10^{15}$ G at the birth. 
The star spins down because of magnetic braking and expels from its superfluid/superconducting
interior magnetic vortices. As a consequence, its magnetic field decreases and 
magnetic reconnections at the star surface produces 
an X-ray luminosity $L_{\mathrm{X,V}}\approx10^{35}\eta\dot{P}^2_{-11}$  erg s$^{-1}$, 
where $\dot{P}_{-11}$ is the period derivative in units of $10^{-11}$ s s$^{-1}$ and $\eta$ 
is an efficiency parameter. In addition, quark star--ring systems can emit X-rays from 
a hot spot formed where steadily accreted ring matter hits the surface.
Transient magnetars are quark star--ring sources that only sporadically enter 
phases of steady accretion. Both shell and ring systems would produce bursts 
when clumps of debris degenerate material are accreted onto the star, the shell 
systems being the most burst-prolific sources, since the co-rotating shell is an 
inherently-less-stable structure than the ring.

In this picture, \sgrze\ is an evolved quark star--ring system that has almost consumed 
its ring, after which it will join other old shell- and ring-less systems with low vortex luminosity 
corresponding, in this scenario, to the XDINSs. The characteristic age of \sgrze\ in 
the quark-nova scenario (where the braking index is $n=4$) is $\tau_c=P/(3\dot{P})\simeq24$ 
Myrs and at this age the magnetic field is expected to have been decayed from the initial value 
of $\sim$$10^{15}$ G to below $\sim$$10^{13}$ G. Reference\,\refcite{ouyed11} shows that 
the rotational and magnetic characteristics of \sgrze\ can account for many of the properties of 
the source observed during the first $\sim$160 days of the 2009 outbursts. 

\section{Conclusion}\label{conclusion}

The discovery of low-$B$ magnetar sources has opened new perspectives in neutron-star 
astrophysics and its consequences deeply impact on our current view of what a `magnetar' 
is. As it was discussed in Sec.\,\ref{theory}, \sgrze\ and \sgrsw\ are likely to be `old magnetars', 
i.e. once-ultra-magnetized neutron stars which experienced substantial field decay over their
(extended, $\approx$1~Myr) lifetime. Still, they retain a large-enough ($\approx$$10^{14}$~G) 
internal toroidal field, sufficient to sporadically produce crustal displacements and hence bursting 
activity. 

In this sense, the original definition of a `magnetar' as a neutron star powered by magnetic 
energy\cite{thompson95} applies to the low-field sources too, at least if one restricts to their active 
phases. What the low-$B$ sources taught us is that this reservoir of magnetic energy (stored in 
the internal field) needs not to show up at all. The external field in a active magnetar can well be 
comparable to that of radio pulsars, dispelling the widespread belief that magnetar activity is necessary 
associated to an ultra-high dipole field. 

The two known low-field sources, \sgrze\ and \sgrsw, are likely not exceptions. Since they represent the `old'
population of initially strongly-magnetized neutron stars, they may constitute the majority of 
magnetar candidates, although their duty cycle is long. Actually, about 20\% of Galactic radio 
pulsars have a dipole field higher that that of \sgrze\ \cite{rea10} and in principle any of them 
may show up as a transient magnetar at any time.

\section*{Acknowledgments}
We are indebted to our too-numerous-to-mention collaborators on works on magnetars, most notably the `Italian magnetar group'. We also wish to thank the organizers and the speakers of the 13th Marcel Grossmann Meeting for the excellent conference. We acknowledge partial funding from INAF through a PRIN-2011 grant. 


\end{document}